# DESIGN OF OPTIMAL REPETITIVE CONTROL BASED ON EID ESTIMATOR WITH ADAPTIVE PERIODIC EVENT-TRIGGERED MECHANISM FOR LINEAR SYSTEMS SUBJECTED TO EXOGENOUS DISTURBANCES

**MOHAMMED SOLIMAN1[1] AND ABDUL-WAHID A. SAIF2[1,2]***

[1]Department of Control and Instrumentation Engineering
[2]Interdisciplinary Research Centre on Smart Mobility and Logistics (IRC-SML)
King Fahd University of Petroleum & Minerals, Dhahran 31261, Saudi Arabia
g202215300@kfupm.edu.sa[1], awsaif@kfupm.edu.sa[2]

## ABSTRACT

The periodic signal tracking and the unknown disturbance rejection under limited communication resources are main important issues in many physical systems and practical applications. The control of such systems has some challenges such as time-varying delay, unknown external disturbances, structure uncertainty, and the heavy communication burden on the sensors and controller. These challenges affect the system performance and may destabilize the system. Hence, in this article, an improved scheme has been designed to overcome these challenges to achieve a good control performance based on optimization technique, and to guarantee the closed-loop system stability. The proposed scheme can be described as: modified repetitive control (MRC) with equivalent-input-disturbance (EID) estimator based on adaptive periodic event-triggered mechanism (APETM). The scheme that has been created is intended for linear systems that experience external disturbances which are not known, and must operate within constraints on communication resources. MRC based on EID has been developed with the goal of achieving periodic reference tracking and enhancing the ability to effectively reject both periodic and aperiodic unknown disturbances. In addition, utilizing APETM to reduce data transmission, computational burden and to save communication resources. Additionally, an optimization method is employed to fine-tune the parameters of the controller, enabling adjustments to the control and learning actions. Overall architecture of the system, incorporating the APETM-MRC with the utilization of an EID estimator and optimal techniques, can be described as a time-varying delay system. Proposed schemes were demonstrated to be effective, feasible, and robust through simulated application.

## 1   INTRODUCTION

The Repetitive Control (RC) algorithm is a highly effective control method that excels in compensating for unknown periodic disturbances and achieving accurate tracking of exogenous periodic signals. It achieves this by employing repeated learning actions, which lead to improved system performance [1], [2]. Due to its ability to generate superior control performance, RC algorithm finds extensive application in various fields. These include servo motors, robotic manipulators, inverted pendulums, and other systems that necessitate the rejection or tracking of periodic signals [3]–[6]. The RC control technique involves leveraging previous errors to adapt the feedback control behavior for the subsequent cycle. This iterative learning process enables a reduction in tracking error and enables accurate tracking of the reference input without steady-state errors. To stabilize a strictly proper plant, the modified RC (MRC) system introduces a Low-Pass Filter (LPF) in addition to the RC system. The integrated LPF serves to reject high-frequency signals, enhancing system stability and significantly reducing control effort. The effects of aperiodic disturbances cannot be rejected with MRC scheme as well as the periodic ones because the RC feedback loop is unable to achieve it [7], and this is a drawback for MRC scheme. In real world, disturbances often consist of a combination of periodic and aperiodic components, occurring at various frequencies. Therefore, a number of control methods based on the MRC methodology have been developed in an effort to get past these disturbances. In order to reject the aperiodic component in addition to the periodic one, equivalent-input-disturbance (EID) based MRC scheme was developed in [8]. Furthermore, numerous control systems using EID-based control strategies have been described, such as [9], [10], due to its efficient performance.

---

* Corresponding Author

Additionally, in older systems, continuous parameters such as outputs, states, or observer states were commonly used. Regardless of whether the data needs updating or not, timing-triggered control regularly performs the necessary updates. This could put a strain on the controller and sensors' communication systems. As a result, the event triggered mechanism (ETM) was widely used in the literature [11], [12]. When compared to other schemes, the ETM provides greater control flexibility and can save more communication resources. ETM can recognize the occurrence of an event to send or not send fresh data over communication channels based on the ET situation. This intended condition is dependent on an error signal that exceeds a predefined limit. The data will not be delivered unless the condition is met, and it will be kept in a zero-order hold (ZOH) [11]. The use of continuous sampling in most ETM approaches can lead to unwanted Zeno behavior due to its dependence on the achieved frequency of conditions [13]. To address this issue, a solution has been proposed in the form of a periodic ETM (PETM) [13]-[15]. The PETM periodically evaluates an ET condition at a predetermined sampling time to determine whether to broadcast a new control signal. In [14], an MRC based on EID was applied to a linear system subjected to both aperiodic and periodic disturbances, utilizing two PETM transmission channels. Excellent tracking performance was achieved by the control scheme, which also rejected aperiodic and periodic disturbances and saved communication resources. Furthermore, When constructing the PETM, a static triggering condition was taken into account. However, this static triggering condition can pose a problem if the dynamics of the system change over time. The static triggering condition lacks the ability to adapt to these changes, potentially leading to suboptimal control performance. Researchers have investigated an adaptive ETM (Event-Triggered Control) technique with a dynamic threshold to preserve computational resources and effectively adapt to alterations in system dynamics [16]-[17]. This article introduces a novel Adaptive PETM (APETM) that combines the advantages of PETM and AETM.

Optimal control is a significant field in modern control theory that aims to design a controller that is better suited to the specific task, thereby enhancing system performance to meet the practical requirements of the real world. The success of this process typically relies on having accurate knowledge of the system's parameters [18]. Optimal RC (ORC) has been applied to a wide range of systems, including robotics, machine tools, and power systems [18]-[20]. This study suggests a design approach for an optimal MRC (OMRC) of a linear system that can simultaneously track a periodic signal and reject external disturbances. OMRC improves upon standard MRC by incorporating an optimization criterion that minimizes a cost function over a finite time horizon. The performance index is used to transform the tracking problem into a Linear Quadratic Regulation (LQR) problem. Tuning the controller parameters affects how well the MRC system performs, in order to determine the optimum state feedback control settings with good optimal solution accuracy and minimal computational complexity, an OMRC algorithm is utilized. The OMRC incorporates state feedback and integral action, which enhances the tracking accuracy and reduces steady-state error, while EID compensation is employed to handle aperiodic disturbances effectively. Overall, OMRC offers improved control performance, robustness, and adaptability to dynamic system changes.

Despite considerable efforts to enhance the control performance of MRC systems, there are still challenges such as time-varying delay, unknown external disturbances, and heavy communication burden on controller and sensors that require further research to develop effective solutions. These challenges can negatively impact the system performance and potentially lead to system instability. Therefore, this paper proposes a combination scheme to address these challenges and improve the MRC system's performance, driven by the following key motivations:

- The proposed control scheme shows superior performance to conventional RC systems in tracking periodic input signals and compensating for both aperiodic and periodic disturbances.
- The APETM is utilized to decrease data transmission, reduce energy use as much as possible and save communication resources.
- A full state observer (FSO) is used to approximate the system states as they cannot be measured directly.
- A state-of-the-art control algorithm is employed to determine optimal state feedback control parameters, achieving high solution accuracy while keeping computational complexity to a minimum.
- To demonstrate its feasibility and effectiveness, the proposed scheme is validated through a simulated application of a speed control of a rotational system with comparative results.

## 2   SYETEM STRUCTURE AND PROBLEM FORMULATION:

In this section, we explore the APETM-OMRC method, which utilizes the EID technique, for linear systems dealing with both unknown periodic and aperiodic disturbances. Figure 1 illustrates the complete system architecture, encompassing the physical plant model with external disturbance, MRC system, FSO, APETM, and EID estimator.

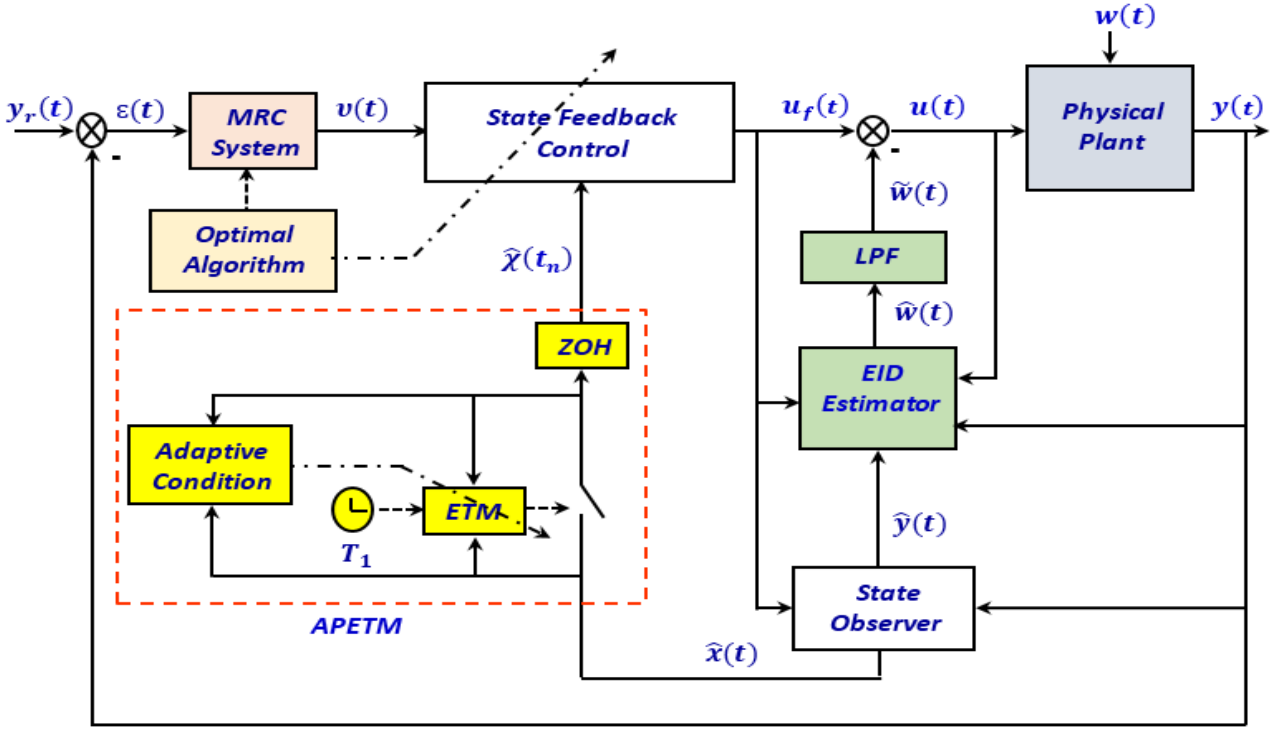

Figure 1: Block diagram of the overall System.

### 2.1   Physical plant model

The given mathematical expression depicts a linear system model with a disturbance input:

$$\dot{x}(t) = Ax(t) + Bu(t) + B_\omega \omega(t)$$

$$y(t) = C x(t) \tag{1}$$

Where, the system is described by state vector $x(t) \in \mathbb{R}^n$, output $y(t) \in \mathbb{R}^p$, control input $u(t) \in \mathbb{R}^m$, and external disturbance input $\omega(t) \in \mathbb{R}^l$. The system is subject to constant matrices $A \in \mathbb{R}^{n \times n}$, $B \in \mathbb{R}^{n \times m}$, $C \in \mathbb{R}^{p \times n}$ and $B_\omega \in \mathbb{R}^{n \times l}$.

## 2.2 MRC System

To track periodic signals and/or reject periodic disturbances, the MRC system is used. In order to streamline the design and execution of the MRC system, the MRC block includes a LPF denoted by $l(s)$, which is positioned before a time delay $e^{-sT}$ in the positive feedback loop, as shown in [7] and as displayed in figure 2.

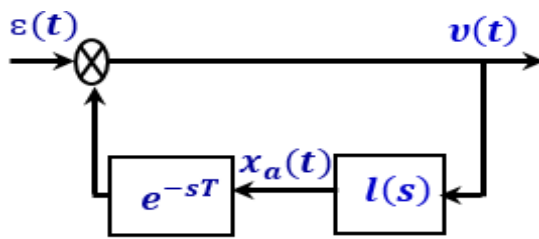

Figure 2: Block diagram of MRC System.

The desired reference signal $y_r(t)$ period $T$, is used to determine the time delay. Specifically, the transfer function of LPF is given by $l(s) = w_a/(s + w_a)$, where the LPF cutoff angular frequency is $w_a$, and $l(s)$ satisfies $|l(jw)| \approx 1$. for all frequencies $w$ in the range $[0, \bar{w}_a]$, where, $\bar{w}_a$ is the maximum angular frequency of the MRC input signal. Therefore, the MRC system's state-space representation looks like this:

$$\dot{x}_a(t) = -w_a x_a(t) + w_a x_a(t-T) + w_a \varepsilon(t)$$
$$v(t) = x_a(t-T) + \varepsilon(t) \qquad (2)$$

Where, $v(t)$ denotes MRC system output, LPF state is $x_a(t)$, and the tracking error, $\varepsilon(t)$ is defined as follows:

$$\varepsilon(t) = y_r(t) - y(t) \qquad (3)$$

## 2.3 FSO

Because system states are assumed immeasurable, they are estimated via an FSO. The system observer dynamics (1) are determined through the following state space:

$$\dot{\hat{x}}(t) = A\hat{x}(t) + Bu_f(t) + L[y(t) - \hat{y}(t)] \qquad (4)$$
$$\hat{y}(t) = C\hat{x}(t)$$

Where, $L \in \mathbb{R}^{n \times p}$ denoted to the designed observer gain, the estimated state vector is $\hat{x}(t) \in \mathbb{R}^n$, and $\hat{y}(t) \in \mathbb{R}^p$ is the estimated output.

### 2.3.1 APETM-FSO based MRC

The first topic to be discussed in this subsection pertains to the APETM-FSO at the observer states channel and then the APETM-FSO based MRC law can be handled. In order to sample the observer states $\hat{x}(t)$ at a predetermined interval $T_1$, an event detector can be used to detect the presence of an event. The event-triggering instants $(t_n)$ can be defined as $t_n = i_n T_1$, where $t_n . T_1 \in \mathbb{N}$, with $i_0 = 0$ and $t_n < t_{n+1}$. In addition, the periodic instants $(l_{n,J})$ can be defined as $l_{n,J} = (i_n + J) T_2, J = 0.1.2.\ldots.d_n$ with $d_n = i_{n+1} - i_n - 1$. Then $[t_n, t_{n+1}) = \bigcup_{J=0}^{t_n} [l_{n,J} . l_{n,J+1})$. The observer error signal $\varepsilon_o(t)$, which is used to create the observer channel ETM condition, is defined as follows:

$$\varepsilon_o(t) = \hat{x}(t_n) - \hat{x}(l_{n,J}) \ . t \in [l_{n,J} . l_{n,J+1}) \qquad (5)$$

Based on $\varepsilon_o(t)$, the event-triggered observer claims that the condition is created as follows:

$$\varepsilon_o^T(t)\psi_1 \varepsilon_o(t) \le \varrho(l_{n,J}) \hat{x}^T(l_{n,J})\psi_2 \hat{x}(l_{n,J}) \qquad (6)$$

Where, $\varrho(l_{n,J})$ is a time-dependent event-triggered threshold function, and $\psi_1, \psi_2$ are the designed symmetric positive definite matrices. The following adaption law can be used to update $\varrho(l_{n,J})$:

$$\varrho(l_{n.J+1}) = Sat_{[\underline{\varrho}.\bar{\varrho}]}\left[\varrho(l_{n.J}) + \kappa\left(\hat{x}^T(t_s)\psi_1\hat{x}(t_s) - \hat{x}^T(l_{n.J})\psi_2\hat{x}(l_{n.J})\right)\right]. \varrho(0) \in [\underline{\varrho}.\bar{\varrho}] \quad (7)$$

Where, $\kappa > 0$ is a parameter to be adjusted.

$$Sat_{[\underline{\varrho}.\bar{\varrho}]}[x] = \begin{cases} \bar{\varrho}. & x \geq \bar{\varrho} \\ x. & \underline{\varrho} \leq x \leq \bar{\varrho} \\ \underline{\varrho}. & x \leq \underline{\varrho} \end{cases} \quad (8)$$

In this context, the parameters and represent the lower and upper bounds of $\varrho(l_{n.J})$. respectively, which correspond to the triggered threshold function. It is important to note that these bounds are subject to the constraint $0 < \underline{\varrho} \leq \bar{\varrho} < 1$. The event-triggered observer defines the channel condition in the following manner: Once condition (6) is violated, an event is triggered, and the current signal is transmitted to the ZOH; at this time, $J = 0$, so $t_n = l_{n.J}$ and $\varepsilon_o(t)$ is reset to zero in accordance with (5); in every other case, the ZOH maintains the previous signal's integrity while the event detector continues to sample the observer state $\hat{x}(t)$., the condition (6) can be verified at the next periodic instance $l_{n.J+1}$. The computation of $i_{n+1}$ is derived as a consequence of this ETM as:

$$i_{n+1} = \min_{J \in \mathbb{N}} \{l_{n.J} | \varepsilon_o^T(t)\psi_1\varepsilon_o(t) > \varrho(l_{n.J}) \ y^T(l_{n.J})\psi_2 y(l_{n.J})\} \quad (9)$$

Therefore, under the planned ETM, the problem of Zeno behavior is resolved. Define $\tau_n(t) = t - l_{n.J}$, $t \in [l_{n.J}, l_{n.J+1})$. Then $\hat{x}(l_{n.J})$ can be written as $\hat{x}(t - \tau_n(t))$. Moreover, from (5), $\hat{x}(t_n)$ can be reformulated as:

$$\hat{x}(t_n) = \varepsilon_o(t) + \hat{x}(t - \tau_n(t)) \quad (10)$$

The APETM-FSO based MRC law is designed as:

$$u_f(t) = k_p\hat{x}(t_n) + k_c v(t) \quad (11)$$

Where. $k_p$ and $k_\varepsilon$ are the intended gains for the controller feedback.

### 2.4 EID Estimator

Incorporating the EID estimator into the existing system facilitates the attenuation of both aperiodic and periodic disturbances. The EID estimation can be expressed as follows, as described in [10]:

$$\hat{\omega}(t) = B^+L[y(t) - \hat{y}(t)] + u_f(t) - u(t) \quad (12)$$

Where, $B^+ = B^T/(B^T B)$. Additionally, output measurement noise could have an impact on the predicted disturbance $\hat{\omega}(t)$, to address this issue, an LPF is used as described in [10]. To facilitate both design and implementation, the LPF can be selected as $f(s) = w_f/(s + w_f)$, with $w_f$ representing the LPF cut off frequency. The state space of LPF can be expressed as:

$$\dot{x}_f(t) = A_f x_f(t) + B_f \hat{\omega}(t) \quad (13)$$
$$\tilde{\omega}(t) = C_f x_f(t)$$

Where, $x_f(t)$, $\tilde{\omega}(t)$ are the state and the output of the filter, respectively. The most recent control law can be obtained by combining the state feedback control law and the output of the EID estimator, expressed as follows:

$$u(t) = u_f(t) - \tilde{\omega}(t) \quad (14)$$

System (1), operating under the enhanced feedback control law (14), achieves successful disturbance attenuation and periodic signal tracking. Additionally, APETM utilization ensures data updates occur only at predetermined intervals, leading to a reduction in communication resources.

### 3 FORMULATION CONTROL DESIGN:

This part commences by providing an overview of the complete system modelling, followed by an analysis of the control design using the proposed technique. Additionally, the final subsection focuses on parameter optimization.

### 3.1 System Modelling

In the proposed design, the external disturbance and periodic reference signal do not impact the stability of the closed-loop system. Consequently, it is adequate to devise a state feedback control rule that remains unaffected by disturbances and periodic references. Since stability is not influenced by external signals, the control design technique simplifies by setting external signals to zero. Consequently, exogenous signals can be considered as zero; $\omega(t) = y_r(t) = 0$, so the system states and tracking error equations can be reformulated as followings:

$$\dot{x}(t) = Ax(t) + Bu(t)$$
$$y(t) = Cx(t) \quad (15)$$
$$\varepsilon(t) = -Cx(t) \quad (16)$$

The equation below captures the estimation error between the observer and system states:

$$x_e(t) = x(t) - \hat{x}(t) \quad (17)$$

As a result, the states $x(t), x_a(t), x_e(t)$, and $x_f(t)$ are used to characterize the augmented system. Moreover, the states equations that describe the overall system are formulated as followings:

$$\dot{x}(t) = (A - Bk_cC)x(t) + Bk_cx_a(t-T) + Bk_Px(t-\tau_n(t)) - Bk_Px_e(t-\tau_n(t)) \quad (18)$$
$$+ Bk_p\varepsilon_o(t) - BC_fx_f(t)$$
$$\dot{x}_e(t) = (A - LC)x_e(t) - BC_fx_f(t) \quad (19)$$
$$\dot{x}_a(t) = -w_ax_a(t) + w_ax_a(t-T) - w_acx(t) \quad (20)$$
$$\dot{x}_f(t) = (A_f + B_fC_f)x_f(t) + B_fB^+LCx_e(t) \quad (21)$$

*Assumption 1:* It is assumed that the system (1) is observable and controlled, that there are no zeros on the imaginary axis, and that flawless performance tracking can be achieved.

### 3.2 Optimization Technique

The following quadratic performance index is introduced in order to derive an ideal controller:

$$J = \lim_{t_\alpha \to \infty} \frac{1}{2} \int_0^{t_\alpha} [z^T(t)Q_z z(t) + u^T(t)Ru(t)]dt \quad (22)$$

Where, $Q_z \in \mathbb{R}^{\bar{n} \times \bar{n}}$ and $R \in \mathbb{R}^{1 \times 1}$ are both positive definite matrices. The following theorem can be used to find the system's ideal control input $u_f(t)$ under the performance index (22).

*Theorem 1:*

Under the performance index (22), the OMRC law $u(t)$ of system (1) with $x(0), y_r(0) = 0$ could be updated as following:

$$u_f(t) = k_p\hat{x}(t_n) + k_cv(t) + f_1(t) \quad (23)$$

$$f_1(t) = -R^{-1}\bar{B}^T \int_0^{l_r} e^{A_c\delta}K\bar{D}r(t+\delta)d\delta \quad (24)$$
$$\qquad -R^{-1}\bar{B}^T \int_0^{l_r} e^{-A_c\delta}K\bar{D}_1\dot{r}(t+\delta)d\delta.$$

$P = [0 \ 0 \ \cdots \ 0 \ 1]^T \in \mathbb{R}^{\bar{n} \times 1}. k_p$ and $k_c$ are given by

$$k_p = -R^{-1}\bar{B}^TK_x, k_c = -R^{-1}\bar{B}^TK_v \quad (25)$$

Where, $K = [K_x \ K_v], K_x \in \mathbb{R}^{\bar{n} \times n}, K_v \in \mathbb{R}^{\bar{n} \times 1}$ is the positive semi-definite matrix satisfying the algebraic Riccati equation (ARE)

$$\bar{A}^TK + K\bar{A} - K\bar{B}R^{-1}\bar{B}^TK + Q_z = 0 \quad (26)$$

Where, $\bar{A} = \begin{bmatrix} A & 0 & 0 \\ -C - CA & -I & 0 \\ -\omega_cC - CA & 0 & -\omega_cI \end{bmatrix}, \bar{B} = \begin{bmatrix} B \\ -CB \\ -CB \end{bmatrix}, \bar{D} = \begin{bmatrix} 0 \\ I \\ \omega_cI \end{bmatrix}, \bar{D}_1 = \begin{bmatrix} 0 \\ I \\ I \end{bmatrix}, \bar{n} = n + 2$

Moreover, $\bar{A}_c$ is defined by: $\bar{A}_c = \bar{A}^T - K\bar{B}R^{-1}\bar{B}^T \quad (27)$

*Design algorithm:*

The following steps can be used to acquire the optimal feedback controller and observer parameters:

**Step 1**: Find the period $T$ of the time delay element that corresponds to the MRC. In addition, for the LPF $l(s)$, determine the cutoff frequency $w_a$ at which the criteria of $l(s)$ are satisfied.
**Step 2**: Choose the matrices $A_f, B_f$ and $C_f$ of so that satisfies the conditions of LPF $f(s)$.
**Step 3:** Specify APETM sample time ($T_1$) and establish the triggered threshold boundaries ($\underline{\varrho}, \bar{\varrho}$).
**Step 4**: Based on the positive matrices $R, Q_z$ and the positive tuning parameters $a, \delta$, Matlab can be used to find a workable solution to the controller update.

## 4 SIMULATION RESULTS:

This section uses the speed control of a rotational system [20] to demonstrate the effectiveness of the proposed OMRC with EID based on APETM design:

$$\begin{cases} \dot{x}(t) = \begin{bmatrix} -31.31 & 0 & -2.83 \times 10^4 \\ 0 & -10.25 & 8001 \\ 1 & -1 & 0 \end{bmatrix} x(t) + \begin{bmatrix} 28.06 \\ 0 \\ 0 \end{bmatrix} u(t). \\ y(t) = [1 \quad 0 \quad 0] x(t). \end{cases} \quad (28)$$

We start by using the most recent control legislation (14) to monitor the subsequent reference input.
$y_r(t) = \sin(\pi t) + 0.5 \sin(2\pi t) + 0.5 \sin(3\pi t)$

The system is subjected to the following periodic and aperiodic disturbances:

$$\omega(t) = \begin{cases} 2 \sin(4\pi t) + \sin(5\pi t) + \sin(6\pi t). & 6 \text{ s} \leq t \leq 8 \text{ s} \\ 3 \sin(2\pi t). & 12 \text{ s} \leq t \leq 18 \text{ s} \\ 0. & \text{otherwise} \end{cases} \quad (29)$$

Moreover, choose $w_a = 100$ rad/sec for MRC filter, and $w_f = 100$ of the EID filter. Consequently, the EID filter's state space parameters are $A_f = -100$, $B_f = 100$, $C_f = 1$. The performance index (22) weight matrices $Q_z, R$, are selected as follows:

$$Q_z = \begin{bmatrix} 100 & 0 & 0 & 0 & 0 \\ 0 & 100 & 0 & 0 & 0 \\ 0 & 0 & 100 & 0 & 0 \\ 0 & 0 & 0 & 20000 & 0 \\ 0 & 0 & 0 & 0 & 0.001 \end{bmatrix}, R = 1$$

Consider $x(0) = [0 \quad 0 \quad 0]^T$ and $u(t) = 0(-2 \text{ s} \leq t \leq 0 \text{ s})$ are the initial conditions. By using the values of other parameters as: $\varphi = \varrho = 0.01$, $T_1 = 0.5$ sec, $\Omega_1 = \Omega_2 = I$, $\psi_1 = \psi_2 = I_{2\times2}$. From theorem (1) by solving the ARE (26) We can get the observer gains and state feedback control, respectively as, $k_p = [-5.0118 \quad 0.1947 \quad 47.4]$, $k_c = 247.25$ and $L = [0.52 \quad 2.215 \quad -0.245]^T$. In the following cases, we go over how the reference and system output signals react.

### 4.1 The proposed controller (APETM-OMRC based on EID) vice MRC scheme case:

In figure 3 with and without an EID estimator, the output $y(t)$ and reference $y_r(t)$ state responses are shown. Without the use of an EID estimator, good tracking performance was achieved in this scenario and the control system can make up for periodic disturbances as shown in figure 3 (a). However, during the 12 – 18 seconds aperiodic disturbance interval, the system cannot compensate for the disturbance as effectively as it does for the periodic disturbance, resulting in a tracking error with an amplitude of 0.15 at time 13.5 sec. as displayed in figure 3 (b). When employing the suggested controller, the output $y(t)$ exhibits minimal tracking error and no overshoot as it closely follows the signal $y_r(t)$. The control system not only compensates for periodic disturbances but also effectively handles aperiodic disturbances. This can be observed in Figure 3, where the maximum tracking error reaches approximately 0.04 at time 1 sec.

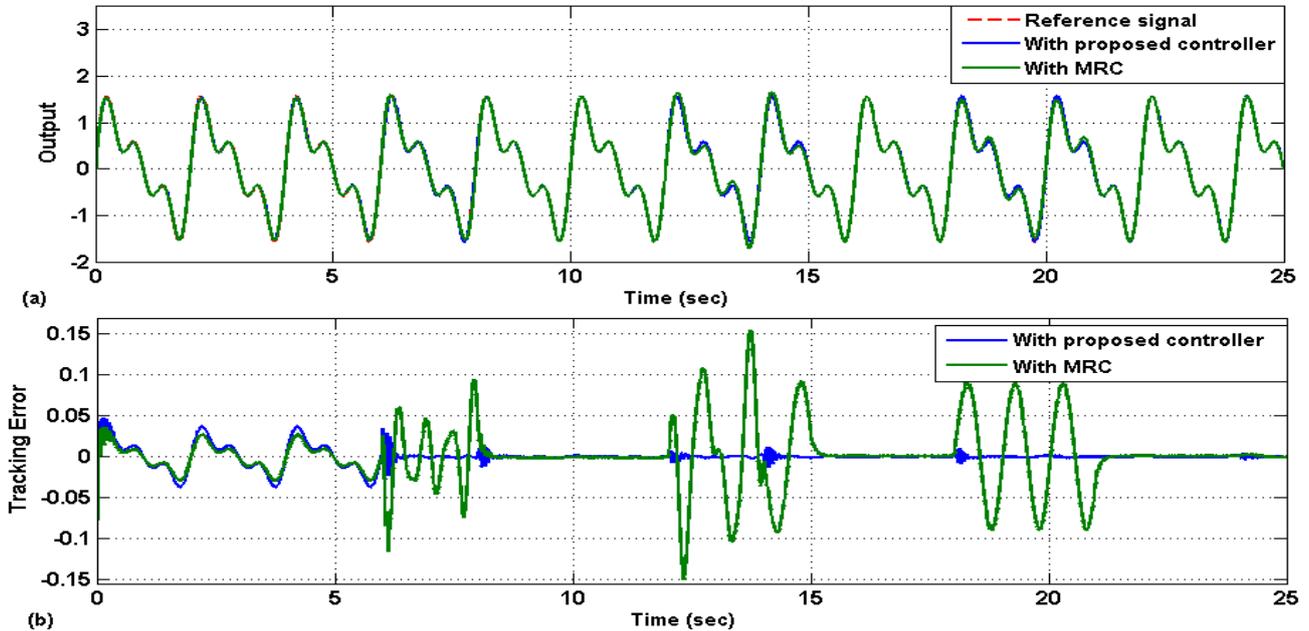

Figure 3: Tracking responses without EID and the proposed controller: (a) $y(t)$. (b) $\varepsilon(t)$.

Moreover, the efficacy of the recommended controller can be observed in the saved communication resources. Figure 4 illustrates the time interval between consecutive event-triggered instances for data transmission. Unlike other control schemes that continuously update control activities, updating control actions only when required helps conserve computational and communication resources.

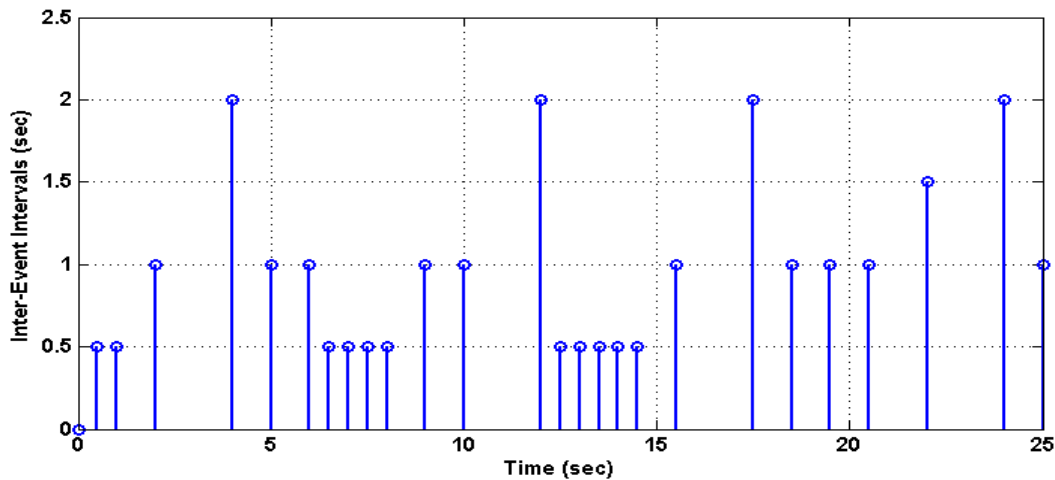

Figure 4: The inter-event intervals under APETM.

As a comparison between the proposed APETM and the PETM scheme which had introduced in [14], [15], the recommended controller's importance can be observed in its capacity to conserve more communication resources. Figure 5 compares the data transmission inter-event intervals of the suggested APETM versus the standard PETM. In comparison to PETM, the proposed APETM exhibits a minimum triggering time that is 20 times longer. Consequently, APETM can be considered more efficient than conventional PETM in terms of reducing data transmission frequency and utilizing fewer communication resources overall.

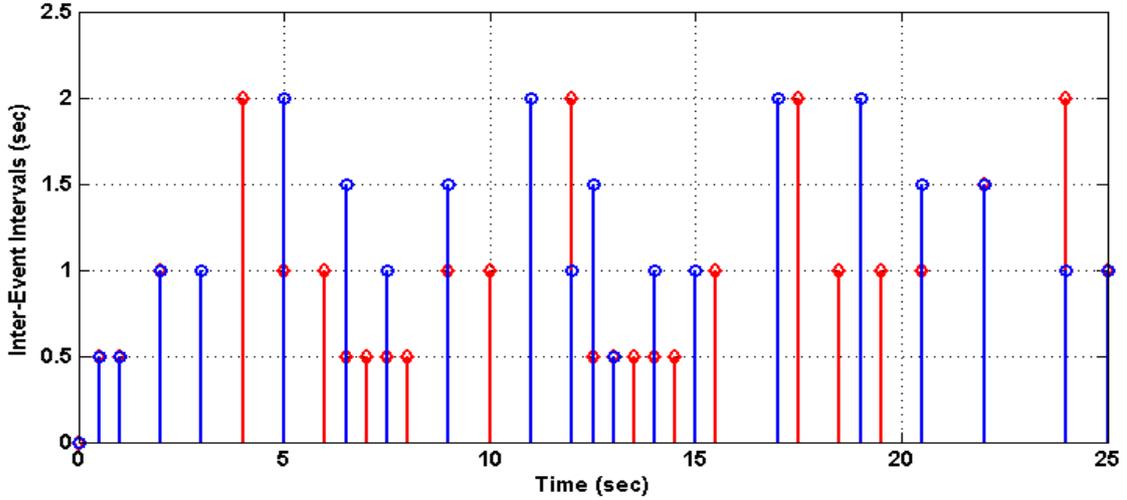

**Figure 5: The inter-event intervals under APETM and PETM in [14] and [15].**

In addition to, three performance indices are used in Table 1 to compare the the proposed controller with the control scheme in [20], which described as: the mean absolute of error (MAE), the mean square of error (MSE), and the root mean square of error (RMSE). These indices are defined as follows:

$$RMSE = \sqrt{\frac{1}{N}\sum_{k=1}^{N}(\varepsilon(k))^2}. \quad MSE = \frac{1}{N}\sum_{k=1}^{N}(\varepsilon(k))^2. \quad MAE = \frac{1}{N}\sum_{k=1}^{N}|\varepsilon(k)|. \quad (30)$$

Where, $N$ is the number of iterations, $y_r(k)$ is the individual data of the output, and $\bar{y}_r$ is the mean value of the output data. Table I demonstrates that the MAE, MSE, and RMSE indices of the proposed controller are lower than those of the other control method in [20]. In light of this, it is possible to conclude that the proposed controller increases system performance in the face of aperiodic and periodic disturbances, as well as time-varying delay.

**Table 1: Comparison Results**

| Comparison | Reference [20] | Proposed controller |
|---|---|---|
| RMSE | 0.3950 | 0.1157 |
| MSE | 0.1561 | 0.0134 |
| MAE | 0.1212 | 0.0295 |

### 4.2 Applying external input disturbance case:

The system's input is disrupted from the outside in order to verify the robustness of the proposed strategy. The proposed controller is capable of compensating for disturbances quickly and achieves good tracking performance, with the maximum tracking error shown in figure 6 (b) being 0.06 when a step signal with amplitude equal to 4.5 is added as an external input disturbance at a time of 21 sec. We started applying a unit step signal and we increased this amplitude gradually and checked the robustness until we reached to this amplitude i.e. 4.5 which is almost 300% of the reference signal. The proposed controller has achieved a good ability to withstand this external disturbance.

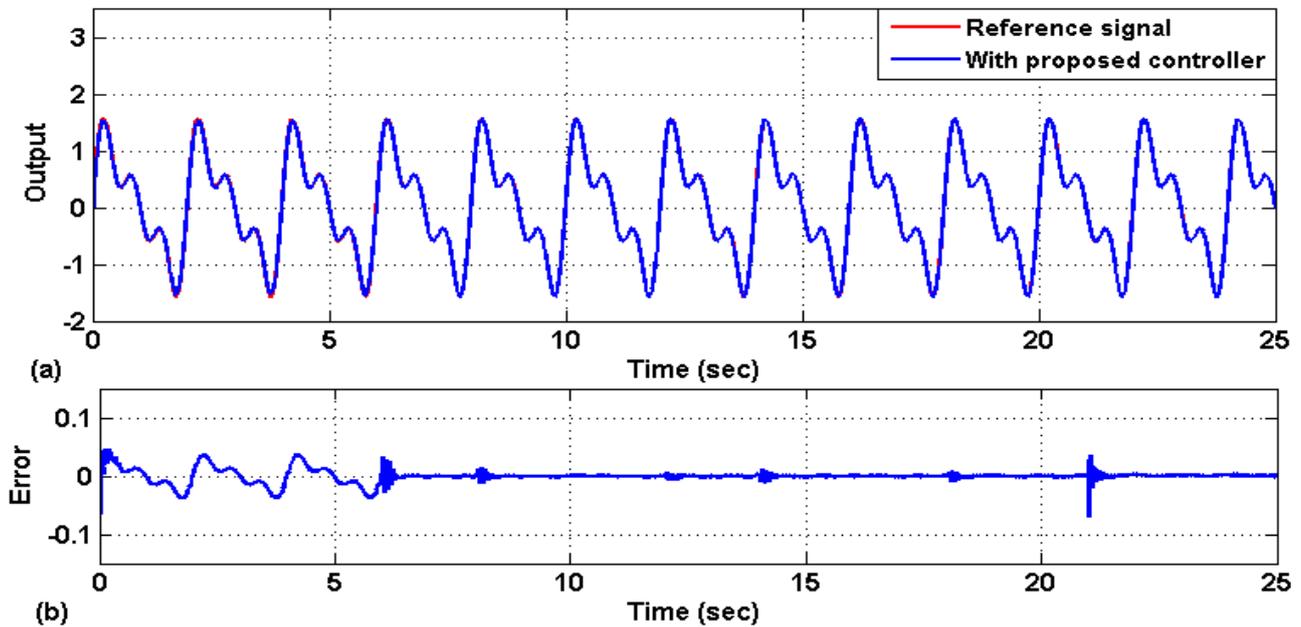

Figure 6: Tracking performance under input disturbance: (a) $y(t)$. (b) $\varepsilon(t)$.

## 5  CONCLUSION:

The objective of this study is to address the challenges associated with tracking control and disturbance rejection in linear systems that are vulnerable to undetected external disturbances. The research proposes a unique technique called EID-based APETM-OMRC, which aims to achieve disturbance rejection, periodic reference tracking, and savings in communication resources. The EID-based MRC design compensates for both periodic and aperiodic disturbances. The proposed controller offers significant value by minimizing communication resources through data transmission only when necessary, thereby reducing energy consumption. Additionally, an adaptive strategy with a shifting threshold has been explored to adapt to changes in system dynamics while conserving computational resources. To optimize the controller parameters and modify the control and learning actions of MRC, an optimal algorithm is utilized. Simulation results from a physical model are provided to demonstrate the superiority and effectiveness of the proposed technique over competing approaches.